\documentclass[preprint,showpacs,preprintnumbers,amsmath,amssymb,11pt]{revtex4}
\usepackage{graphicx,natbib,color,cancel,soul}
\usepackage{dcolumn}
\usepackage[utf8]{inputenc}
\usepackage{bm}

\newcommand{\beq}{\begin{eqnarray*}}
\newcommand{\eeq}{\end{eqnarray*}}
\begin{document}

\title{Observational constraints on a cosmological model with Lagrange multipliers}

\author{Antonella Cid}
\email{acidm@ubiobio.cl}
\affiliation{Departamento de F\'{\i}sica, Universidad del B\'io-B\'io, Avenida Collao 1202, Casilla 5-C, Concepci\'{o}n, Chile}

\author{Pedro Labra\~na}
\email{plabrana@ubiobio.cl}
\affiliation{Departamento de F\'{\i}sica, Universidad del B\'io-B\'io, Avenida Collao 1202, Casilla 5-C, Concepci\'{o}n, Chile}

\begin{abstract}
Cosmological models with Lagrange multipliers are appealing because they could explain the behaviour of the dark sector in a unified way. 
In this work we analyse extensions to the ``Dust of Dark Energy model'' proposed in \cite{Lim} by including spatial curvature and more general potentials of the scalar field. We perform dynamical system analysis and we determine the evolution of the equation of state parameter as a function of the scale factor. We present observational constraints on this model by using Union2.1 dataset and H(z) data.
\end{abstract}

\maketitle
\section{Introduction}

In 1998 measurements of the luminosity distance of supernovae type Ia (SnIa) indicated the unexpected result that the Universe is undergoing accelerated expansion~\cite{Super}, which would be driven by a negative-pressure matter component called dark energy. 

On the other hand, astrophysical observations provide compelling evidence \cite{DM} for the existence of a non-baryonic, non-interacting and pressure-less component of the Universe, dubbed dark matter. This component clusters allowing structures to form.

The existence of both dark components is supported by observations such as cosmic microwave background (CMB) \cite{WMAP7}, baryon acoustic oscillations (BAO) \cite{Cluster}, Hubble constant measurements \cite{Riess}, SnIa \cite{Super,Suzuki}. The available data indicate that 72.9$\%$ of the total matter content is dark energy, $22.6\%$ is dark matter and the remaining $4.5\%$ corresponds to baryonic matter \cite{WMAP7}.

Given that we can not measure direct evidence of dark matter or dark energy, there exist a degeneracy in the dark sector \cite{degeneracy}. This degeneracy allow us to explore different dark candidates for the matter content of our universe. In this sense, it is very appealing to consider a single fluid describing the dark sectors in a unified way, which behaves as dark matter in early epochs and as a mixture of dark matter and dark energy nowadays. The archetypical unified model is the Chaplygin Gas \cite{Chaplygin0}, which has been widely studied in several context \cite{Chaplygin1}.

Recently, a new model for unifying the dark sector has been proposed \cite{Lim}. This model, named Dust of Dark Energy (DDE), describes the dark sector by using two scalar fields where one of them is a Lagrange multiplier which imposes a constraint on the dynamics. In this sense, the dark sector is described by a single fluid which could represent dust or dark energy in different epochs of the evolution.
The DDE model is appealing because it could be consistent with structure formation as suggested in Ref.\cite{Lim}. Cosmological models with Lagrange multipliers (LM) has been studied in different context \cite{DDE}-\cite{DDE4}. For example, in \cite{DDE1} the role of LM was analyse in the context of $f(R)$ gravity, in \cite{DDE2} the Hamiltonian formalism was developed in modified gravity, in \cite{DDE3} the authors investigate cyclic and singularity free scenarios in the context of modified gravity with LM and in \cite{DDE4} cosmological models with LM were studied with the focus on the cosmological constant value.

On the other hand, the more recent cosmological data seem to favour a slightly closed geometry for our universe. A joint analysis with CMB, BAO and SnIa indicate: $\Omega_{k_0}=-0.0057^{+0.0067}_{-0.0068}$ \cite{WMAP7}.

In this work we extend the study of the model developed in \cite{Lim} by considering: more general potentials for the scalar field and spatial curvature. We perform dynamical system analysis and put constraints on the parameters of the model by  using the Union 2.1 sample of SnIa and the expansion rate data H(z). This paper is structured as follows: in section \ref{I} we describe the model, in section \ref{DSA} we use dynamical system analysis in order to study the asymptotic behaviour of the model. In section \ref{NS} we show the numerical solution to the differential equations describing the model, in section \ref{sec5} we perform Bayesian analysis with supernovae and H(z) data. Finally, in section \ref{cl} we resume our results.

\section{The model}
\label{I}
The model is described by the action \cite{Lim}:
\beq
S=\int d^4x\sqrt{-g}\left[-\frac{R}{2\kappa^2}+K(\psi,X)+\lambda\left(X-\frac{1}{2}\mu^2(\psi)\right)\right]
\eeq
where $K$ is a function of the scalar field $\psi$ and the kinetic term $X=\frac{1}{2}g^{\alpha\beta}\nabla_{\alpha}\psi\nabla_{\beta}\psi$. $\lambda$ is a Lagrange multiplier, $\mu^2$ is the potential of the scalar field $\psi$, $R$ is the Ricci scalar, $g$ is the metric determinant, $\kappa^2$ is a normalization constant and we consider $c=1$. From this action we get the following set of field equations:
\beq
G_{\mu}^{\nu}=\kappa^2T_{\mu}^{\nu},\ \ \ \ \ \ \nabla_{\nu}T_{\mu}^{\nu}=0\ \ \ \ \ \textrm{and}\ \ \ \ \ X=\frac{1}{2}\mu^2(\psi)
\eeq
where $G_{\mu}^{\nu}$ is the Einstein tensor and $T_{\mu}^{\nu}=\textrm{diag}(\rho,-p,-p,-p)$ is a perfect fluid-type energy-momentum tensor. The total energy density $\rho$ and the total pressure $p$ are defined respectively given by \cite{Lim}:
\begin{eqnarray}
\label{M1}
\rho=\mu^2(K_X+\lambda)-K\ \ \ \ \ \ \ \ \ \ \textrm{and}\ \ \ \ \ \ \ \ \ \ p= K
\end{eqnarray}
In order to get the explicit form of the field equations we consider the Friedmann-Robertson-Walker metric in co-moving coordinates with a non zero curvature term:
\begin{eqnarray*}
ds^2=dt^2-a^2(t)\left(\frac{dr^2}{1-kr^2}+r^2d\theta^2+r^2\sin^2\theta d\phi^2\right)
\end{eqnarray*}
where $a(t)$ is the scale factor and the curvature parameter $k=0,+1,-1$ represents flat, closed and open spatial sections, respectively. 

By imposing homogeneity and isotropy to the field equations we can consider $X=\frac{1}{2}\dot{\psi}^2$ and, because of the constraint, $\mu=\dot{\psi}$. Dots denote derivatives respect to the cosmological time.

By choosing $K=-X$ we recover the dynamics of the dark sector, where $\rho=\frac{\mu^2}{2}\left(2\lambda-1\right)$, $p=-\frac{\mu^2}{2}$ and the derivative of the state parameter $\omega=\frac{p}{\rho}$ turns to be:
\begin{eqnarray}
\label{eq1}
\omega'=\frac{2\lambda'}{\left(2\lambda-1 \right)^{2}}=2\omega^2\lambda',
\end{eqnarray}
Here primes denote a derivative respect to $\log a$. We note that in order to have $\rho>0$ we need to fulfil the condition $\lambda>\frac{1}{2}$ or equivalently $\omega<0$, which will be assumed from now on. We recover a cosmological constant-type fluid for a constant potential $\mu^2=\mu_0^2$ and $\lambda=1$, given that in this case $\rho=\frac{\mu_0^2}{2}$ and $\omega=-1$.

By combining the conservation equation, $\rho'+3(\rho+p)=0$, the Friedmann equation, $H^2+\frac{k}{a^2}=\frac{\rho}{3}$ and the definition of $\rho$ in Eq.(\ref{M1}) we get:
\begin{eqnarray}
\label{eq2}
\lambda'=(1-2\lambda)\left[\frac{3}{2}(1+\omega)-\epsilon\sqrt{\frac{-6\omega}{1-k\chi}}\right]
\end{eqnarray}
where we have defined $\epsilon=-\frac{\mu_{\psi}}{\mu}$ and $\chi=\frac{3}{\rho a^2}$. $\mu_{\psi}$ denotes the derivative of $\mu$ respect to the scalar field $\psi$ and $H=\frac{\dot{a}}{a}$ is the expansion rate. We have used $\kappa=1$.

The dynamical set of equations describing the model is conveniently chosen to be given in terms of the functions $\omega$, $\chi$ and $\epsilon$ as:
\begin{eqnarray}
\label{eq3}
\omega'&=&3\omega\left[(1+\omega)-\frac{2}{3}\epsilon\sqrt{\frac{-6\omega}{1-k\chi}}\right]\\
\label{eq4}
\chi'&=&\chi(1+3\omega)\\
\label{eq5}
\epsilon'&=&-\epsilon^2(\Gamma-1)\sqrt{\frac{-6\omega}{1-k\chi}}
\end{eqnarray}
where we defined $\Gamma=\frac{\mu_{\psi\psi}\mu}{\mu_{\psi}^2}$. We note that by assuming $\Gamma$ as a function of $\epsilon$ it is possible to get a closed set of equations. By providing $\Gamma(\epsilon)$, the function $\mu(\psi)$ is determined by the solution of the following differential equation, $\Gamma(\epsilon)\mu_{\psi}^2=\mu_{\psi\psi}\mu$, where we have to use the definition of $\epsilon$ in terms of the potential $\mu$. In this sense, to provide $\Gamma(\epsilon)$ is equivalent to define the scalar field potential $\mu$. See TABLE \ref{T3} for simple examples.

\section{Dynamical System Analysis}
\label{DSA}
\subsection{Constant $\epsilon$}
In order to study the set of Eqs.(\ref{eq3})-(\ref{eq5}) we begin by considering the simplest case of a constant $\epsilon$. There are two possibilities to get a constant $\epsilon$: a potential $\mu(\psi) = Cte.$ which implies $\epsilon = 0$ or $\mu(\psi) = A e^{B\psi}$, for a non-zero constant $\epsilon$. In both cases the Eq.(\ref{eq5}) is trivially satisfied.

Under these considerations the dynamical set of Eqs.(\ref{eq3})-(\ref{eq5}) is reduced to a two dimensional system described by:
\begin{eqnarray}\label{dins1}
\omega' &=& 3\, \omega \left[1 + \omega -\sqrt{\frac{\omega}{\omega_f}}\,\frac{(1 + \omega_f)}{\sqrt{1 - k\,\chi}}\,\right],\\
\chi' &=& (1 + 3\, \omega)\,\chi\,,\label{dins2}
\end{eqnarray}
where we have conveniently define $\omega_f$ by $\epsilon = \frac{3}{2\sqrt{6}}\frac{1+\omega_f}{\sqrt{-\omega_f}}$, see \cite{Lim}. From this definition we note that for a constant potential $\omega_f=-1$, otherwise $\omega_f$ will be a negative constant in order to have a real valued $\epsilon$.

\begin{figure}[ht!]
\begin{center}
\includegraphics[width=2.5in,angle=0,clip=true]{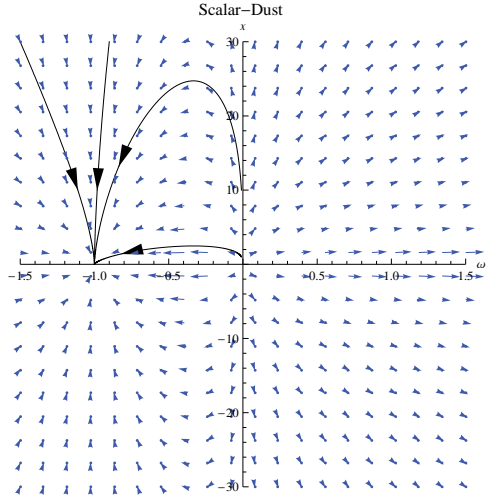}
\caption{\label{Nsol1}The Direction Field of Eqs.~(\ref{dins1})-(\ref{dins2}) and the numerical solution for 4 different sets of initial conditions. Case $\omega_f=-1$.} 
\end{center}
\end{figure}

\begin{figure}[ht!]
\centering
\includegraphics[width=6cm,angle=0,clip=true]{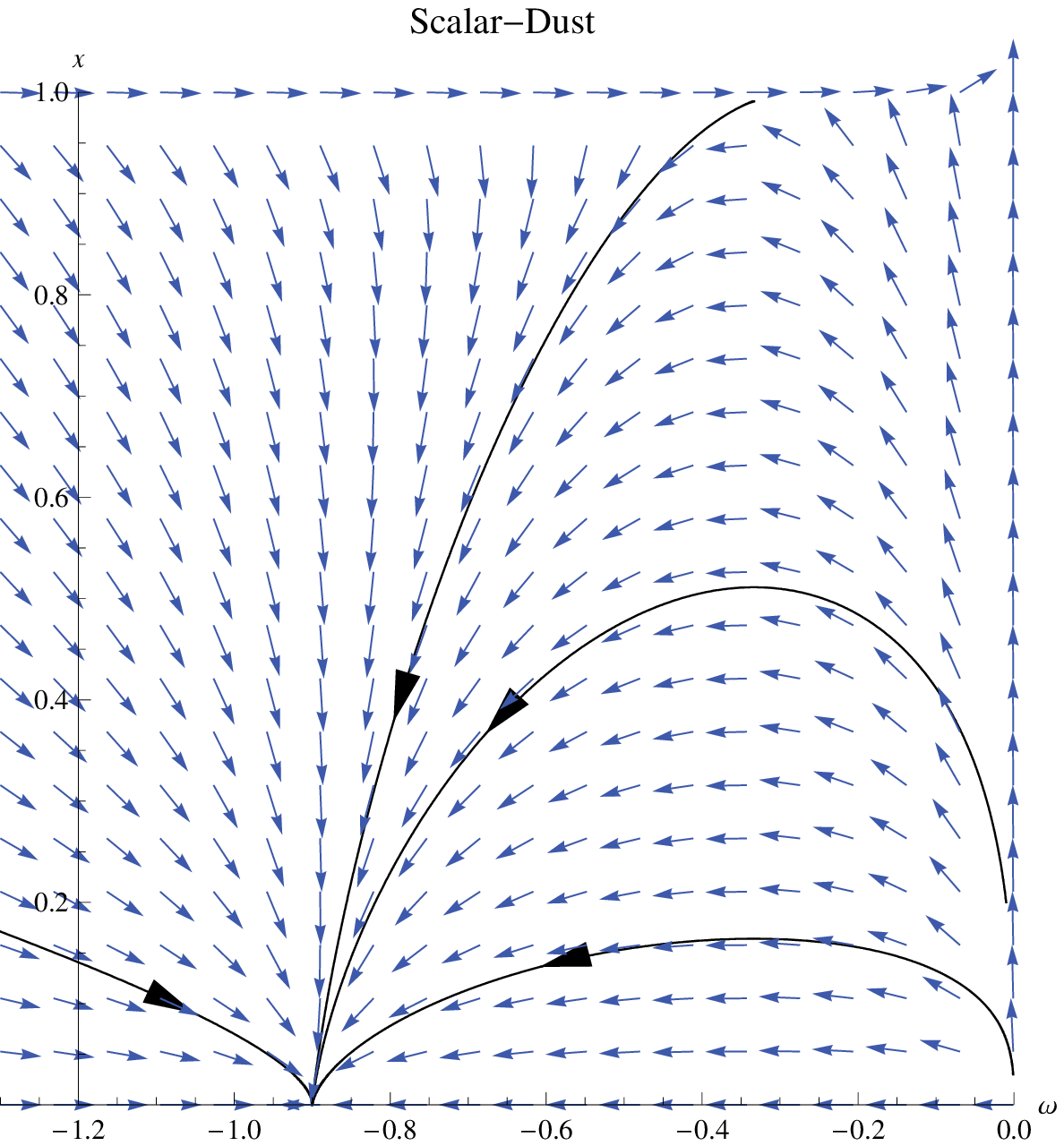}
\includegraphics[width=6cm,angle=0,clip=true]{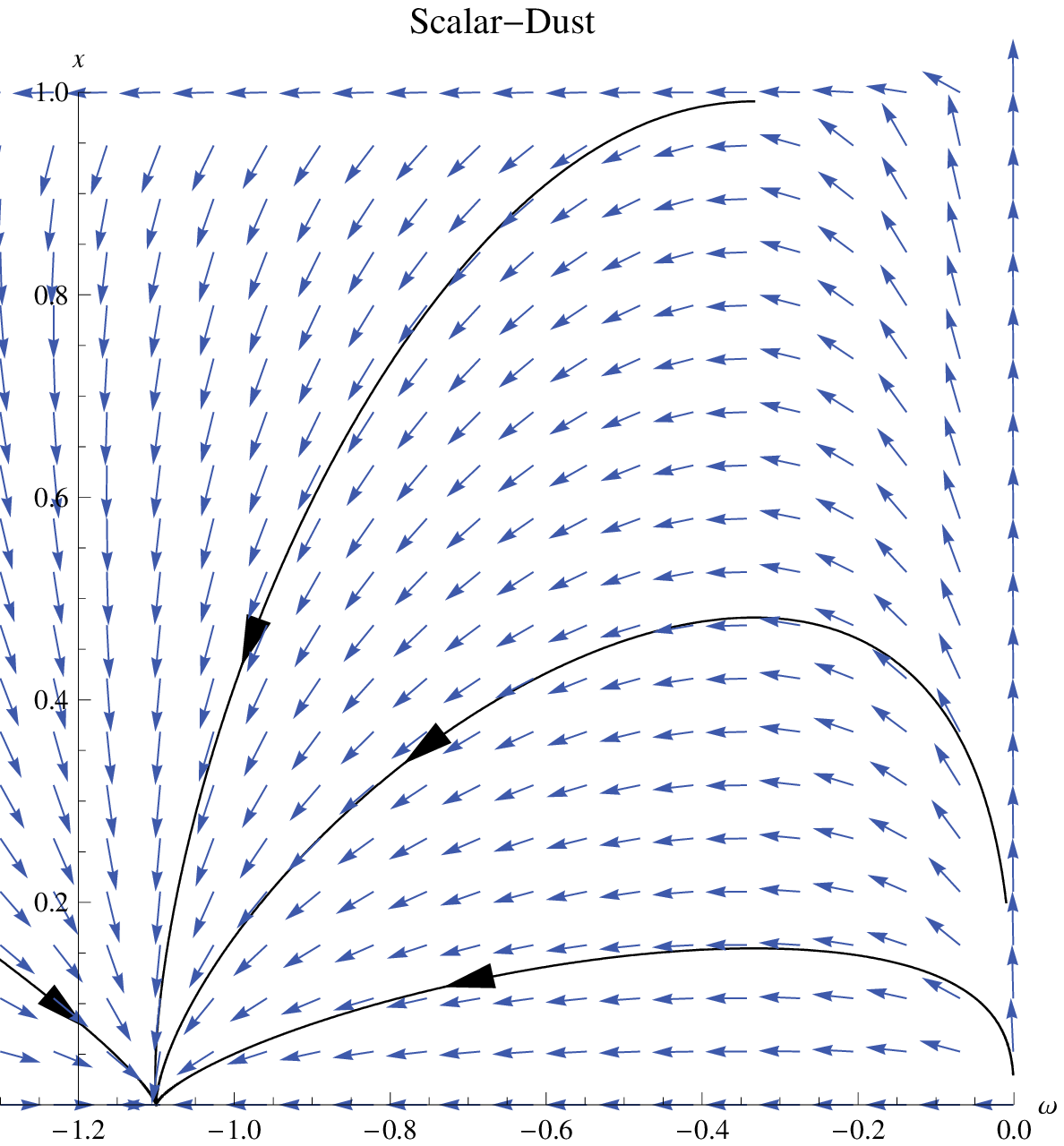}
\caption{\label{Nsol2} The Direction Field of Eqs.~(\ref{dins1})-(\ref{dins2}) and the numerical solution for 4 different sets of initial conditions. Case $k=1$, left panel $\omega_f=-0.9$ and right panel $\omega_f=-1.1$}
\end{figure}

\begin{figure}[ht!]
\centering
\includegraphics[width=6cm,angle=0,clip=true]{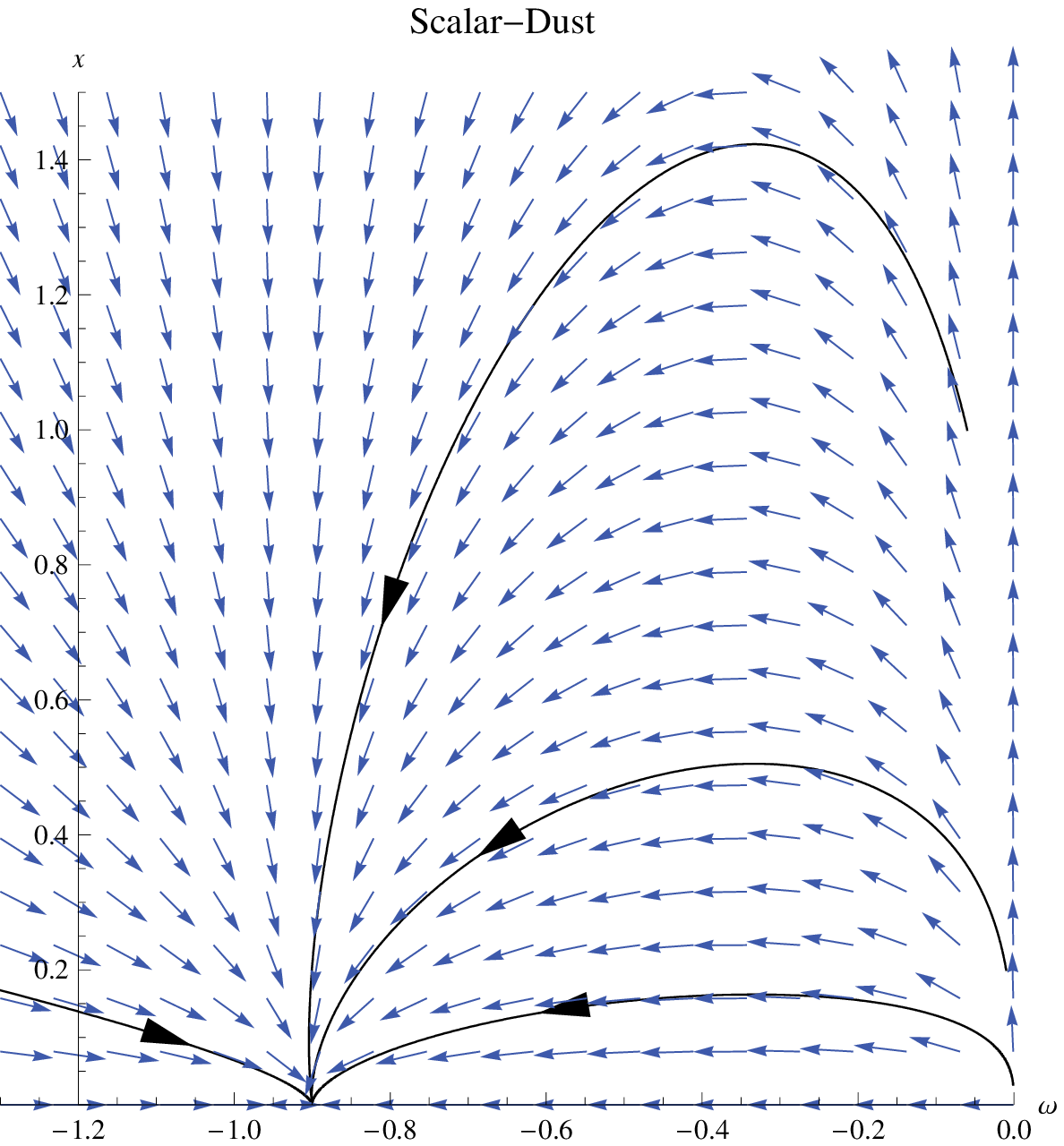}
\includegraphics[width=6cm,angle=0,clip=true]{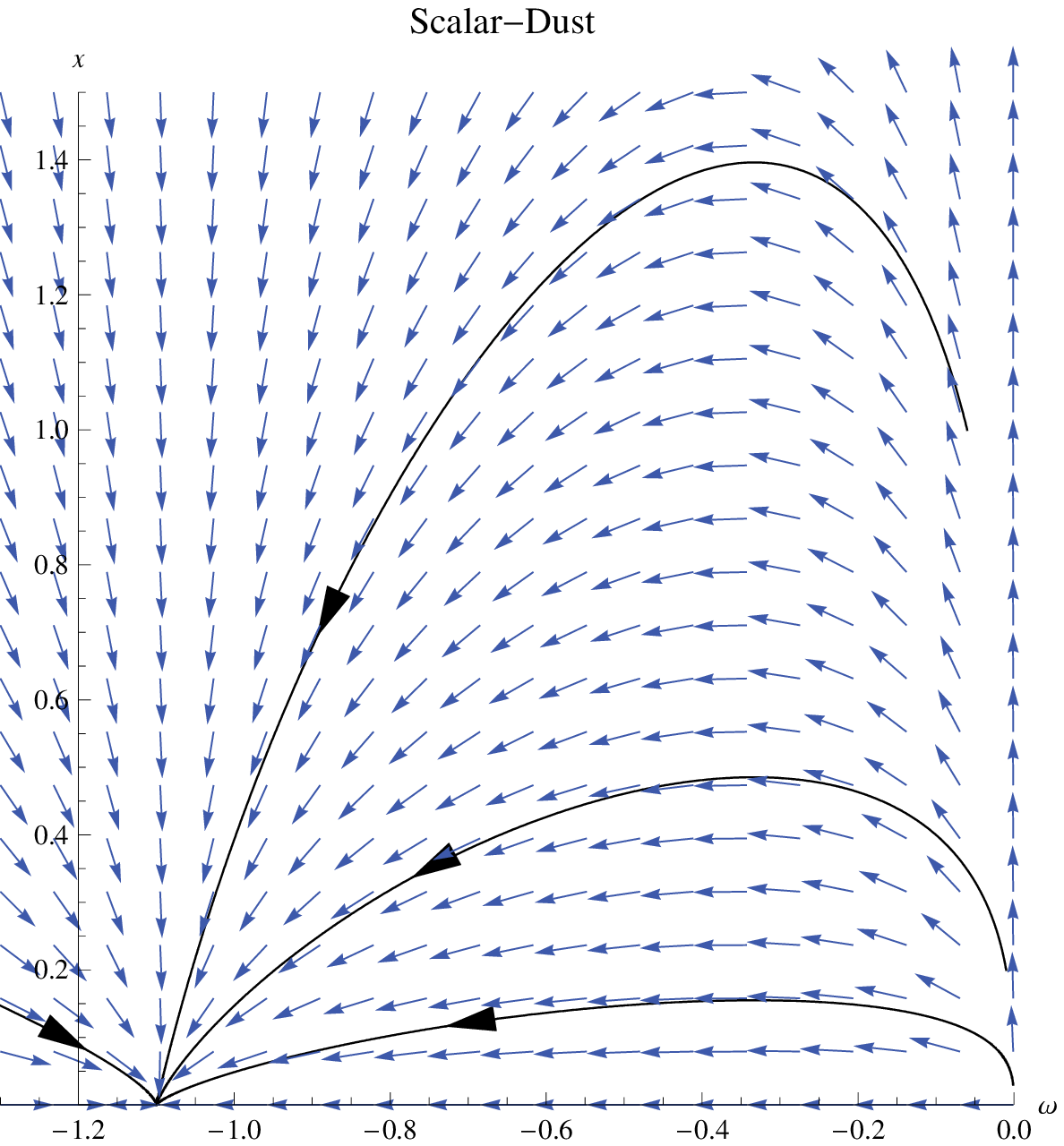}
\caption{\label{Nsol3} The Direction Field of Eqs.~(\ref{dins1})-(\ref{dins2}) and the numerical solution for 4 different sets of initial conditions. Case $k=-1$, left panel $\omega_f=-0.9$ and right panel $\omega_f=-1.1$}
\end{figure}

The critical points of the system and their main characteristics are given in TABLE \ref{T1} where the conditions for the existence of the critical points are shown.
The most interesting critical points are 1 and 2 which could be consider as the past and future evolution of the universe, respectively. In this case the universe evolves from a state dominated by a fluid with $\omega = 0$ (dust) to a state dominated by a fluid with $\omega = \omega_f < -1/3$ (dark-energy). This behaviour is corroborated by the numerical integration of Eqs.(\ref{dins1})-(\ref{dins2}), see FIGs. \ref{Nsol1}-\ref{Nsol3}.

\begin{table}[ht!]
\centering
\begin{tabular}{|c|c|c|c|c|c|}\hline
N&$\chi_c$ & $\omega_c$ & Stability & Condition& Curvature\\ \hline
1&0 & 0 & unstable node & No & any\\ \hline 2&0 & $\omega_f$ &
attractor & $\omega_f<-\frac{1}{3}$& any\\ \hline 3&0 & $\omega_f$ &
unstable node & $-\frac{1}{3}<\omega_f<0$& any\\ \hline
4&$-\frac{10}{4}-\frac{3}{4\omega_f}-\frac{3\omega_f}{4}$ & $-\frac{1}{3}$ & center& $-\frac{1}{3}<\omega_f<0$ & $k=-1$\\ \hline
5&$\frac{10}{4}+\frac{3}{4\omega_f}+\frac{3\omega_f}{4}$ & $-\frac{1}{3}$ &
saddle point & $-1<\omega_f<-\frac{1}{3}$ & $k=1$\\ \hline
\end{tabular}
\caption{\label{T1} Critical points and stability behaviour for the system of Eqs.(\ref{dins1})-(\ref{dins2}). We have considered $\chi>0$ and $\omega_f < 0$, which follow from the definition of these variables.} 
\end{table}

As an example, we show in FIGs. \ref{Nsol1}-\ref{Nsol3} the phase space for four numerical solution to Eqs.(\ref{dins1})-(\ref{dins2}) and different values of the parameters $\omega_f$ and $k$. In these figures we have included the Direction Field of the system in order to have a picture of whatever a general solution looks like. In particular, in FIG. \ref{Nsol1} it is shown the case where $\omega_f=-1$. In this case the curvature term is irrelevant, as we note from Eqs.(\ref{dins1})-(\ref{dins2}), and we reproduce the result in \cite{Lim}. Note that in the figure the physical part of the plot is delimited by $\omega \leq 0$ and $\chi \geq0$. In FIG. \ref{Nsol2} the cases $k=1$ for, $\omega_f =-0.9$ and $\omega_f=-1.1$ are shown. In FIG. \ref{Nsol3} the cases $k=-1$ for $\omega_f =-0.9$ and $\omega_f=-1.1$  are shown.

It is interesting to note that for a nearly flat scalar potential where $\epsilon\ll1$ and approximately constant, the system of Eqs.(\ref{eq3})-(\ref{eq5}) can be reduced to a two dimensional system \cite{delCampo1}. The critical points consistent with this kind of potential are 1 and 2 of TABLE \ref{T1}, because in this case $\omega_f$ has to be close to $-1$.

In section \ref{sec5} we are going to contrast this particular model to cosmological observations via Bayesian methods, which will allow us to constraint the parameters of the model.
\subsection{Variable $\epsilon$}
In order to study more general behaviour for the solution of Eqs.(\ref{eq3})-(\ref{eq5}), where $\epsilon$ is not a constant, we consider a family of potentials (see TABLE \ref{T3}) which generate a simple structure for the $(\Gamma-1)$ term in Eq.(\ref{eq5}) as $(\Gamma-1)\propto\epsilon^n$ where $n$ is an integer and $n\ge-1$.

This family of potentials allows that the set of Eqs.(\ref{eq3})-(\ref{eq5}) becomes a three dimensional autonomous system with a critical point for $\epsilon=0$, $\chi=0$ and $\omega=0$ or $\omega=-1$.
\begin{table}[ht!]
\begin{tabular}{l l l l}\hline\hline
$\Gamma-1=-n^{-1}$       & $\rightarrow$ &\ \ \ \  $\mu(\psi)=\mu_0\psi^n$           & \\ 
$\Gamma-1=\epsilon^{-1}$ & $\rightarrow$ &\ \ \ \  $\mu(\psi)=C_1 e^{-C_2e^{-\psi}}$ & \\ 
$\Gamma-1=C \epsilon^n$  & $\rightarrow$ &\ \ \ \  $\mu(\psi)=C_1 e^{-\frac{((n+1)(\psi+C_2))^{n/(n+1)}}{n}}$, n even\\ 
$\Gamma-1=C \epsilon^n$  & $\rightarrow$ &\ \ \ \  $\mu(\psi)=C_1 e^{\pm\frac{((n+1)(\psi-C_2))^{n/(n+1)}}{n}}$, n odd\\\hline\hline
\end{tabular}
\caption{\label{T3}The family of potentials which close the autonomous system of Eqs.(\ref{eq3})-(\ref{eq5})}
\end{table}

\begin{figure}
\centering
\includegraphics[width=6cm]{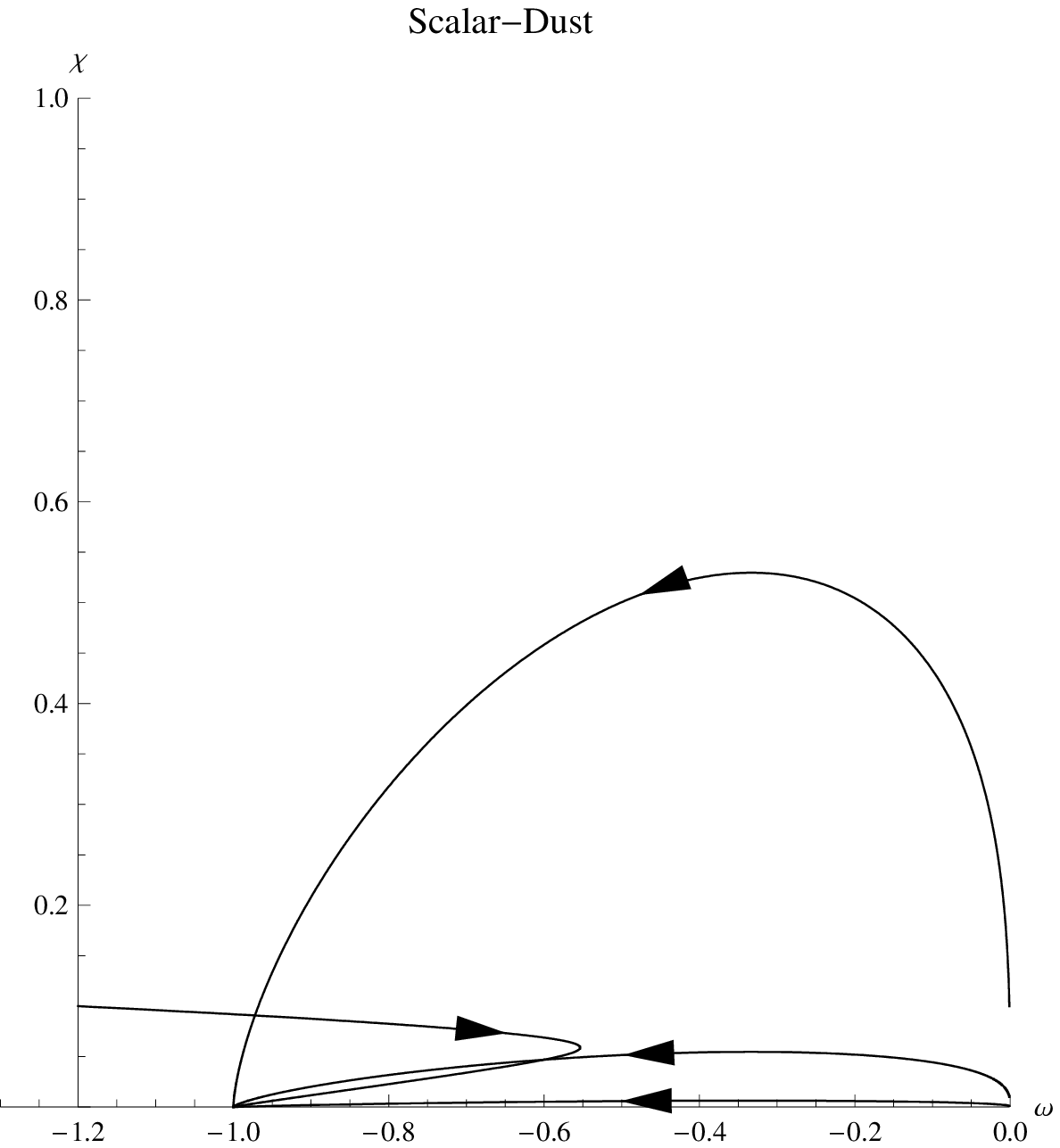}
\includegraphics[width=6cm]{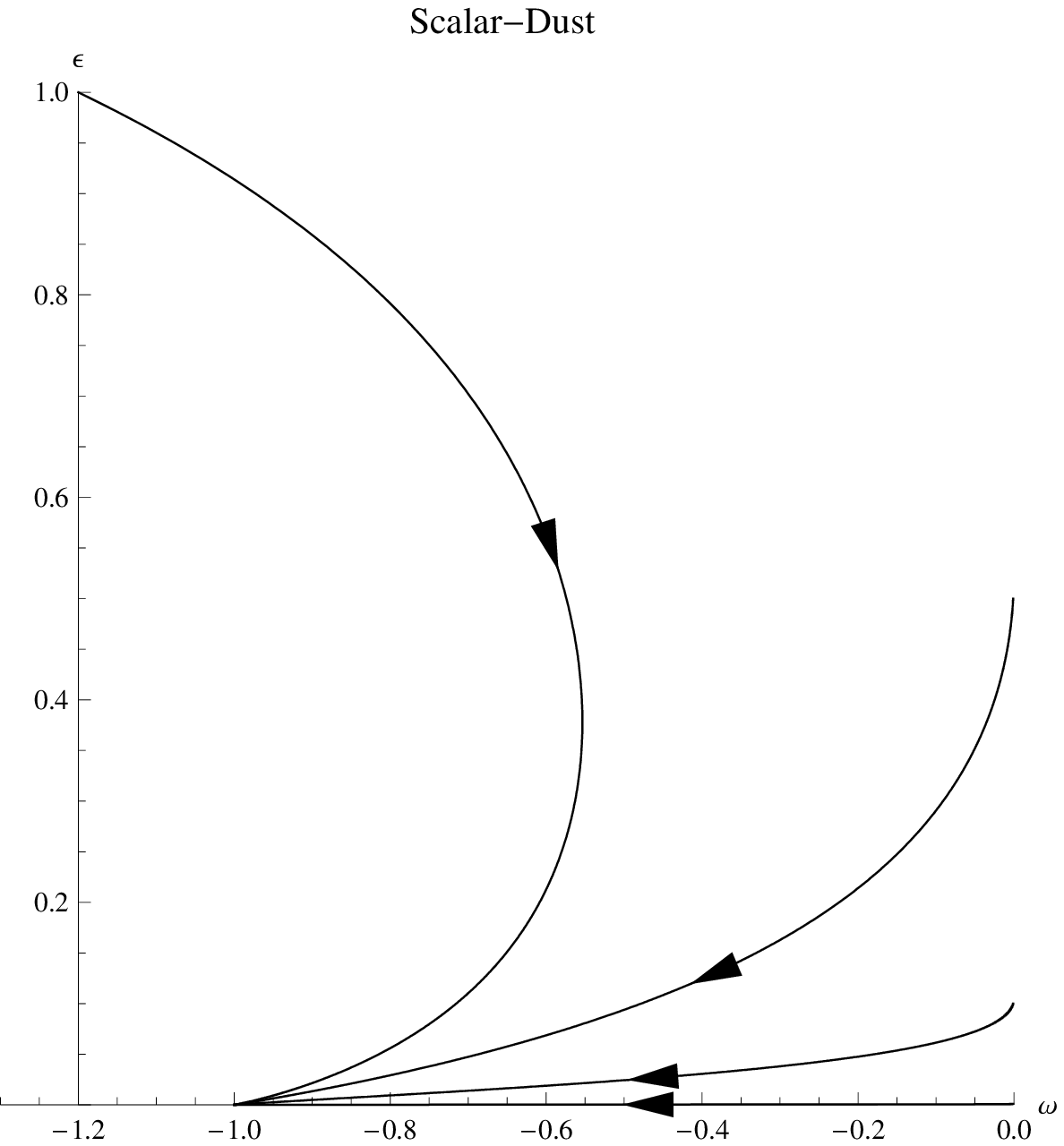}
\caption{\label{Fig-epsilon}Some numerical solutions for Eqs.(\ref{ec-epsilon1})-(\ref{ec-epsilon3}) where we have consider as an example $k=-1$ and different initial conditions. The left panel is a projection to the axis $(\omega, \chi)$ and the right panel is a projection to the axis $(\omega,\epsilon)$.}
\end{figure}

As an example, let us show the case of a potential such that $\Gamma-1=\epsilon^{-1}$. In this case, the dynamical set of equations becomes:
\begin{eqnarray}
\omega' &=& 3\, \omega \left[1 + \omega -\frac{2}{3}\,\epsilon\,\sqrt{\frac{-6\,\omega}{1 - k\,\chi}}\,\right],\label{ec-epsilon1}\\ 
\chi' &=& (1 + 3\, \omega)\,\chi\,,\label{ec-epsilon2}\\
\epsilon' &=& - \epsilon\,\sqrt{\frac{-6\,\omega}{1 -k\,\chi}}.\label{ec-epsilon3}
\end{eqnarray}
This system have two critical points given in Table \ref{T2}. The critical point $1^*$ is an unstable focus and the critical point $2^*$ is an attractor. Similar to the case discussed above, we can consider these critical points as the past and future evolution of the universe and both are independent of the curvature. In this case the universe evolves from a state dominated by a fluid with $\omega = 0$ to a state dominated by a fluid with equation the state $\omega = -1$ (cosmological constant). This behaviour is corroborated by the numerical integration of Eqs.(\ref{ec-epsilon1})-(\ref{ec-epsilon3}). Some of these solution are given in FIG.~\ref{Fig-epsilon}, where a projection of the solution to the axes $(\omega, \chi)$ and $(\omega, \epsilon)$ is shown, for several choices of the initial conditions.

\begin{table}[ht!]
\centering
\begin{tabular}{|c|c|c|c|c|c|c|}\hline
N&$\chi_c$ & $\omega_c$ & $\epsilon_c$ & Stability & Condition& Curvature\\
\hline $1^*$& $0$ & $0$ & $0$ & unstable node & No & any\\ \hline
$2^*$ &
$0$ & $-1$ & $0$ & attractor & No & any\\
\hline
\end{tabular}
\caption{\label{T2} Critical points and stability behaviour for the system of Eqs.(\ref{ec-epsilon1})-(\ref{ec-epsilon3}).}
\end{table}

\section{Numerical Solution for a nearly flat scalar potential}
\label{NS}
In order to numerically integrate the set of Eqs.(\ref{dins1})-(\ref{dins2}) we use more convenient functions defined as $\Omega=\frac{\rho}{3H^2}$ and $\Omega_k=-\frac{k}{a^2H^2}$ at any time in the evolution. In terms of these new variables Eqs.(\ref{dins1})-(\ref{dins2}) are transformed to:
\begin{eqnarray}
\label{eq6}
\omega'&=&3\omega\left[(1+\omega)-\sqrt{\frac{\omega}{\omega_f}}(1+\omega_f)\sqrt{\Omega}\right]\\
\label{eq7}
\Omega'&=&\Omega(\Omega-1)(1+3\omega)
\end{eqnarray}

\begin{figure}[ht!]
\centering
\includegraphics[width=0.45\columnwidth]{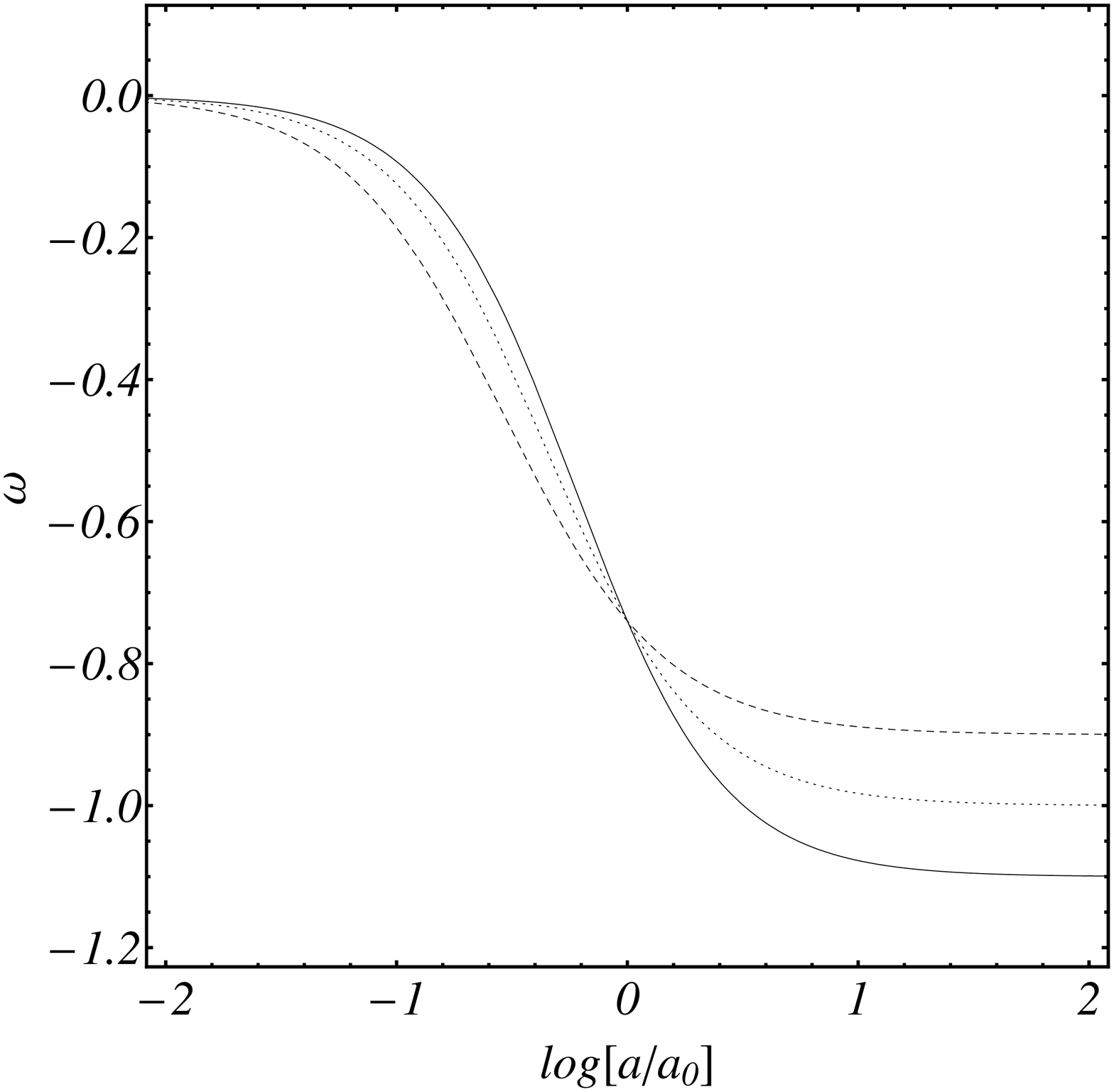}
\includegraphics[width=0.485\columnwidth]{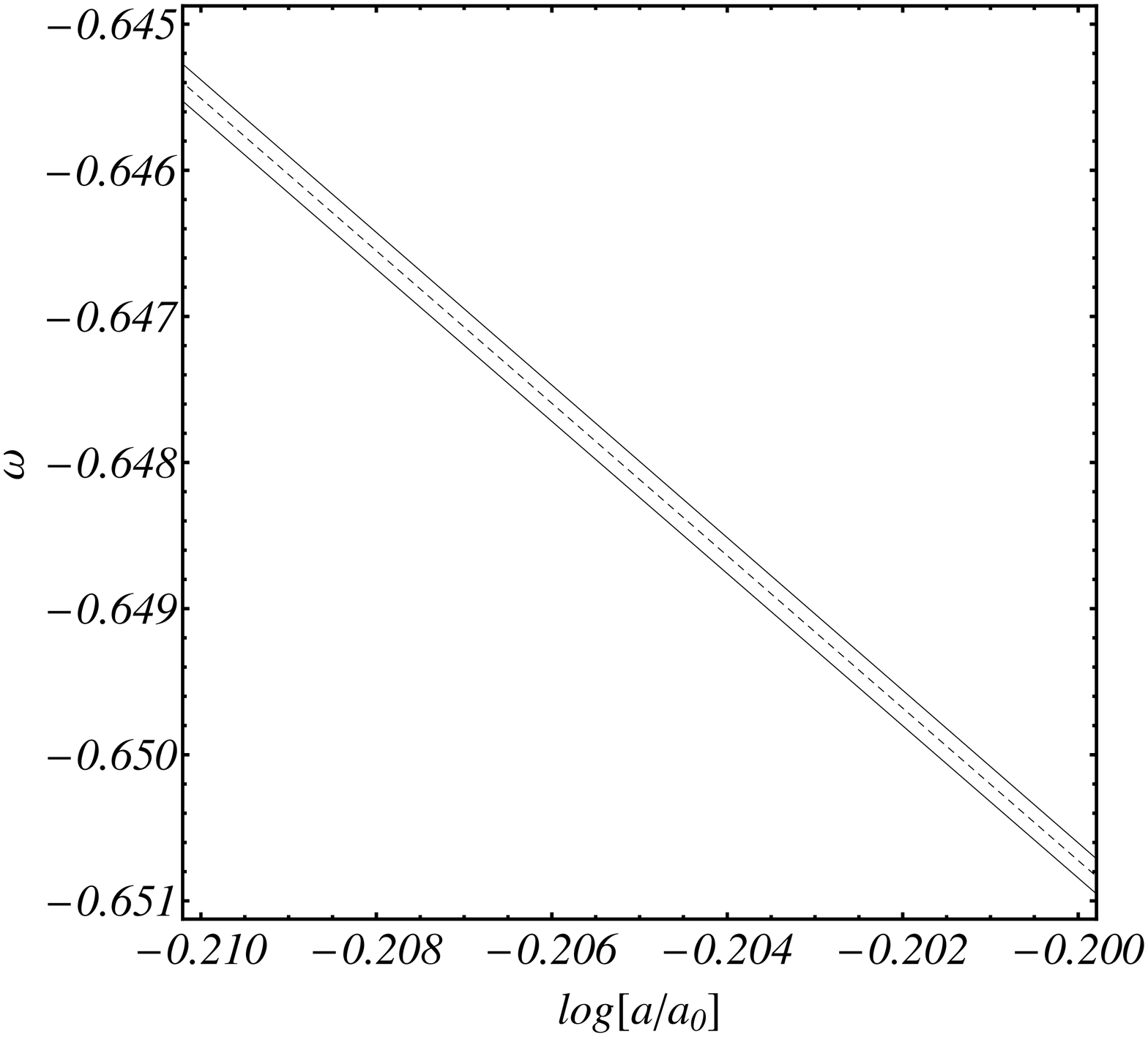}
\caption{\label{mono2} The left panel corresponds to the evolution of state parameter for a flat spatial section. We have used $\omega(0)=\omega_0=-0.74$ which is consistent with the cosmological data that we present in section \ref{sec5}. The continuous line corresponds to $\omega_f=-1.1$, the dotted and dashed lines are for $\omega_f=-1$ and $\omega_f=-0.9$ respectively. The right panel shows a closed region for $\omega_f=-0.9$ and $\Omega_{k_0}=+0.005,0,-0.005$ from top to bottom.} 
\end{figure}

By considering contributions to the spatial curvature of order $\Omega_{k_0}=\pm0.005$ (consistent with the data \cite{Suzuki}) there are no significant modifications in the evolution of $\omega(a)$ 
as it is shown in FIG. \ref{mono2}. Here the subscript 0 denotes the value of a function today.

In FIG. \ref{mono2} we see that the fluid behaved like dust in the past ($a<a_0$), with a state parameter close to zero. This behaviour is independent of the allowed value of $\omega_f$ and consistent with small contributions of spatial curvature, as we noted in the dynamical system analysis. In the future ($a>a_0$) the state parameter reaches a constant value corresponding to $\omega=\omega_f$, which is also independent of the curvature. When $\omega_f=-1$ the fluid asymptotically becomes a cosmological constant, whereas for $\omega_f<-1$ the fluid asymptotically becomes a phantom fluid.

\begin{figure}[ht!]
\centering
\includegraphics[width=0.471\columnwidth]{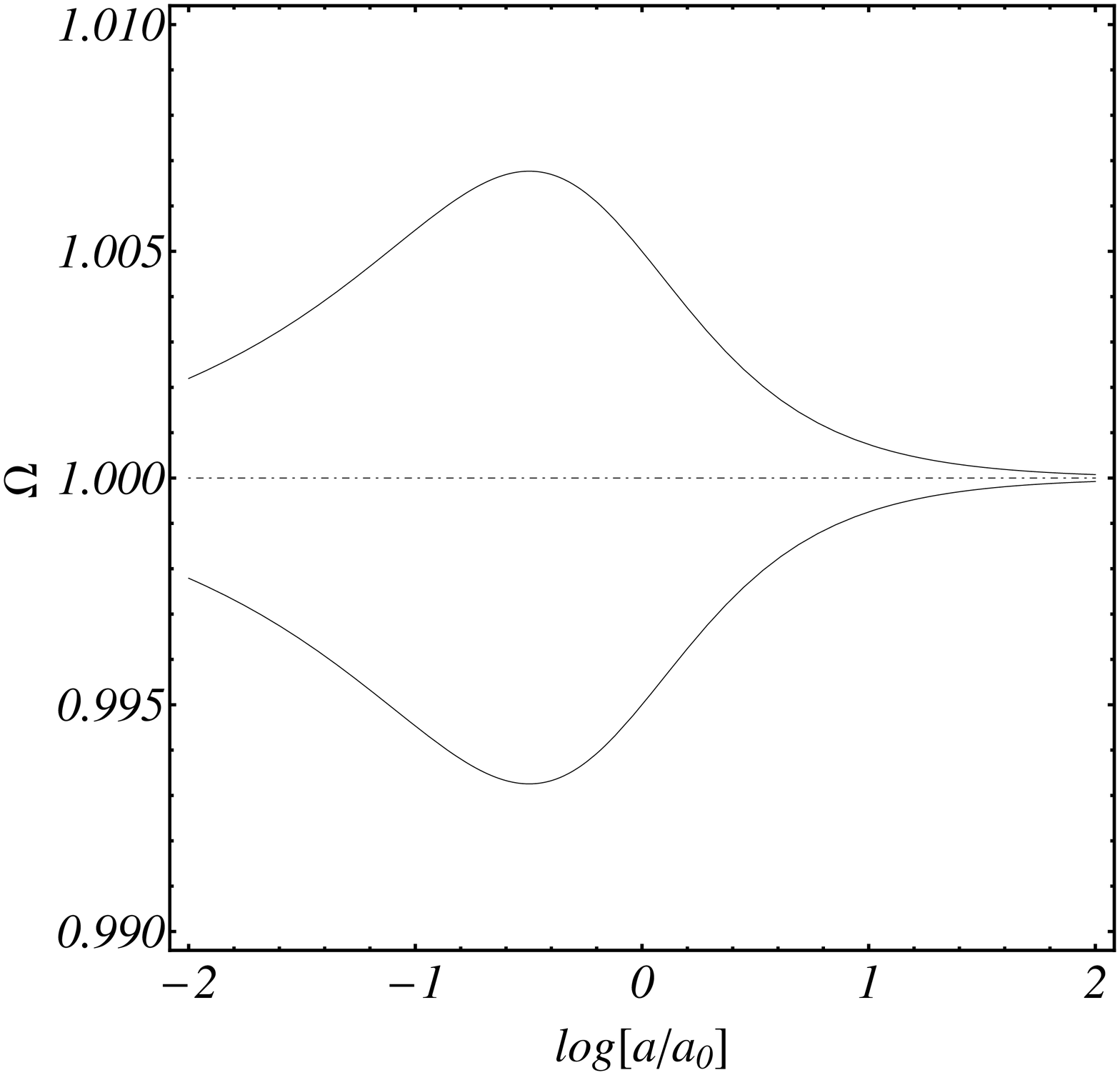}
\includegraphics[width=0.45\columnwidth]{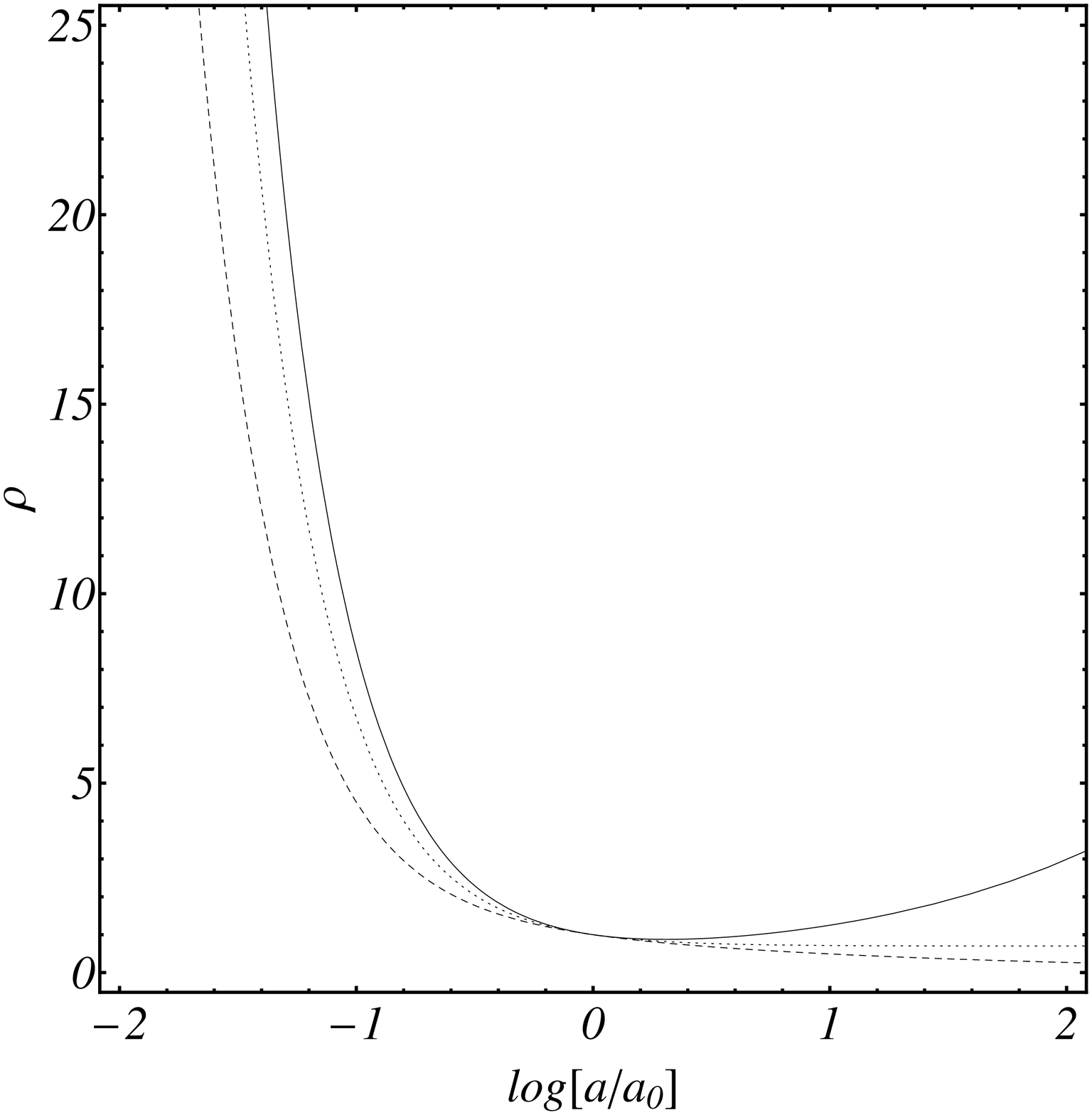}
\caption{\label{mono3} The left panel shows the evolution of the density function $\Omega$ for three different values of $\Omega_{k_0}$. The continuous lines correspond to $\omega_0=-0.74$ and $\omega_f=-1.1$, for $\Omega_{k_0}=\pm0.005$. The dotdashed line is for $\Omega_{k_0}=0$. The right panel shows the evolution of the energy density with $\omega_f=-1.3,-1,-0.8$ for solid, dotted and dashed lines, respectively. We normalized to have $\rho=1$ today and we have considered $\omega_0=-0.74$.}
\end{figure}

The curves in the left panel of FIG. \ref{mono3} are the result of numerical integration of Eqs.(\ref{eq6})-(\ref{eq7}) with $\omega(0)=\omega_0=-0.74$ for different values of $\Omega_{k_0}$. The area between the solid lines expands a continuos range of values of the curvature parameter, inside the current observational constraints \cite{Suzuki}. As we noted in the dynamical system analysis, the effect of curvature is not significant in the initial or final state of the universe in the context of this model.

The right panel of FIG. \ref{mono3} shows the numerical integration of the energy density $\rho$. For $\omega_f=-1.1$ and $\omega_0=-0.74$, the final state of our universe will be dominated by a cosmological constant fluid. For $-\frac{1}{3}<\omega_f<-1$ the energy density goes to zero at the end and for $\omega_f<-1$ we have a universe which will be dominated by a phantom fluid at the end, where the energy density $\rho$ increases in the future  without bound. 

\section{Observational Data Analysis}
\label{sec5}
In this section we examine the observational constraints on the model defined by Eqs.(\ref{dins1})-(\ref{dins2}), with and without spatial curvature. We use SnIa observations and H(z) data. 

We perform bayesian statistical analysis using SnIa data from the Supernova Cosmology Project Union2.1 sample \cite{Suzuki}, with 580 supernovae over the range $0.015<z<1.414$.

We fit the theoretical distance modulus $\tilde{\mu}_{th}(z)$ defined by:
\beq
\tilde{\mu}_{th}(z)=5\log_{10}\left[\frac{H_0d_L(z)}{c}\right]+\tilde{\mu}_0
\eeq
to the corresponding observed distance modulus $\tilde{\mu}_{obs,i}$. Here $\tilde{\mu}_0=42.38-5\log_{10}h$, $H_0=100h$ [kms$^{-1}$Mpc$^{-1}$] is the Hubble constant and the luminosity distance is defined as $d_L(z)=(1+z)r(z)$ with \cite{Weinberg}:
\beq
r(z)=\frac{c}{H_0\sqrt{\vert\Omega_{k_0}}\vert}S_k\left[\sqrt{\vert\Omega_{k_0}\vert}\int_0^z\frac{dz}{H(z)}\right],
\eeq
for $S_k(x)=\sin(x),x,\sinh(x)$ for $k>0,k=0,k<0$, respectively.

The constraints from the SnIa data can be obtained by minimizing the following $\chi^2$ function:
\beq
\chi^2_{\tilde{\mu}}(\tilde{\mu}_0,{\bf p})=\sum_{i=1}^{580}\left(\frac{\tilde{\mu}_{obs,i}-\tilde{\mu}_{th}(z_i;\tilde{\mu}_0;{\bf p})}{\sigma_{\tilde{\mu}}(z_i)}\right)^2
\eeq
where ${\bf p}$ represents the model parameters and $\sigma_{\tilde{\mu}}(z_i)$ is the distance-modulus uncertainty for the corresponding redshift $z_i$.

It is not difficult to realize that $\tilde{\mu}_0$ is a nuisance parameter and we can easily marginalize over it \cite{Li:2010du}. Thus instead of minimizing $\chi^2_{\tilde{\mu}}$ we minimize the function $\tilde{\chi}_{\tilde{\mu}}^2$ which is independent of the $\tilde{\mu}_0$ parameter.
\beq
\tilde{\chi}_{\tilde{\mu}}^2({\bf p})=A({\bf p})-\frac{B({\bf p})^2}{C({\bf p})}, \ \textrm{where}
\eeq
\beq
A({\bf p})=\sum_{i=1}^{580}\left(\frac{\tilde{\mu}_{obs,i}-\tilde{\mu}_{th}(z_i;\tilde{\mu}_0=0;{\bf p})}{\sigma_{\tilde{\mu}}(z_i)}\right)^2;\ 
B({\bf p})=\sum_{i=1}^{580}\frac{\tilde{\mu}_{obs,i}-\tilde{\mu}_{th}(z_i;\tilde{\mu}_0=0;{\bf p})}{\sigma_{\tilde{\mu}}(z_i)^2};\
C({\bf p})=\sum_{i=1}^{580}\frac{1}{\sigma_{\tilde{\mu}}(z_i)^2}
\eeq

\begin{figure}[ht!]
\centering
\includegraphics[width=0.47\columnwidth]{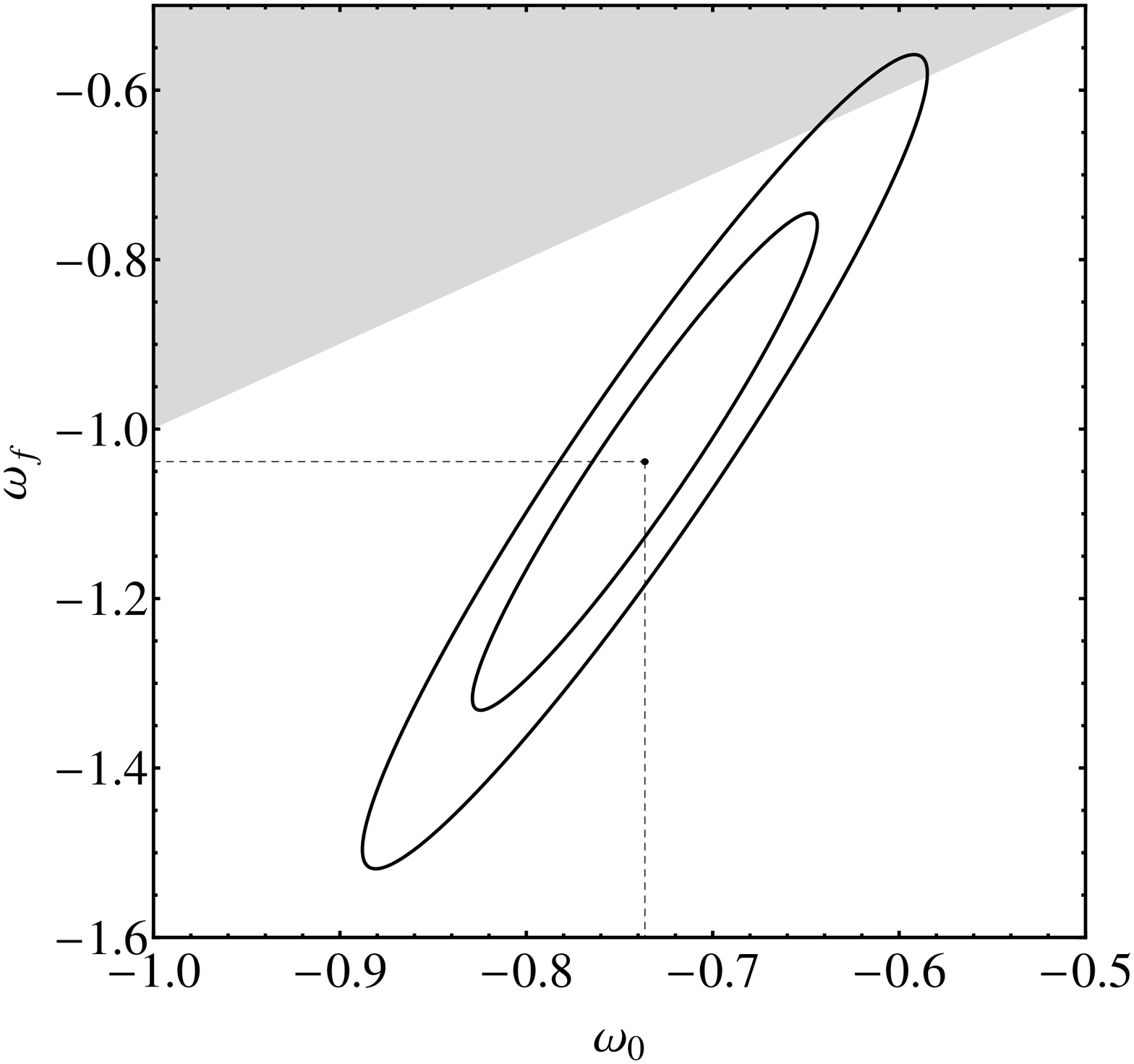}
\includegraphics[width=0.45\columnwidth]{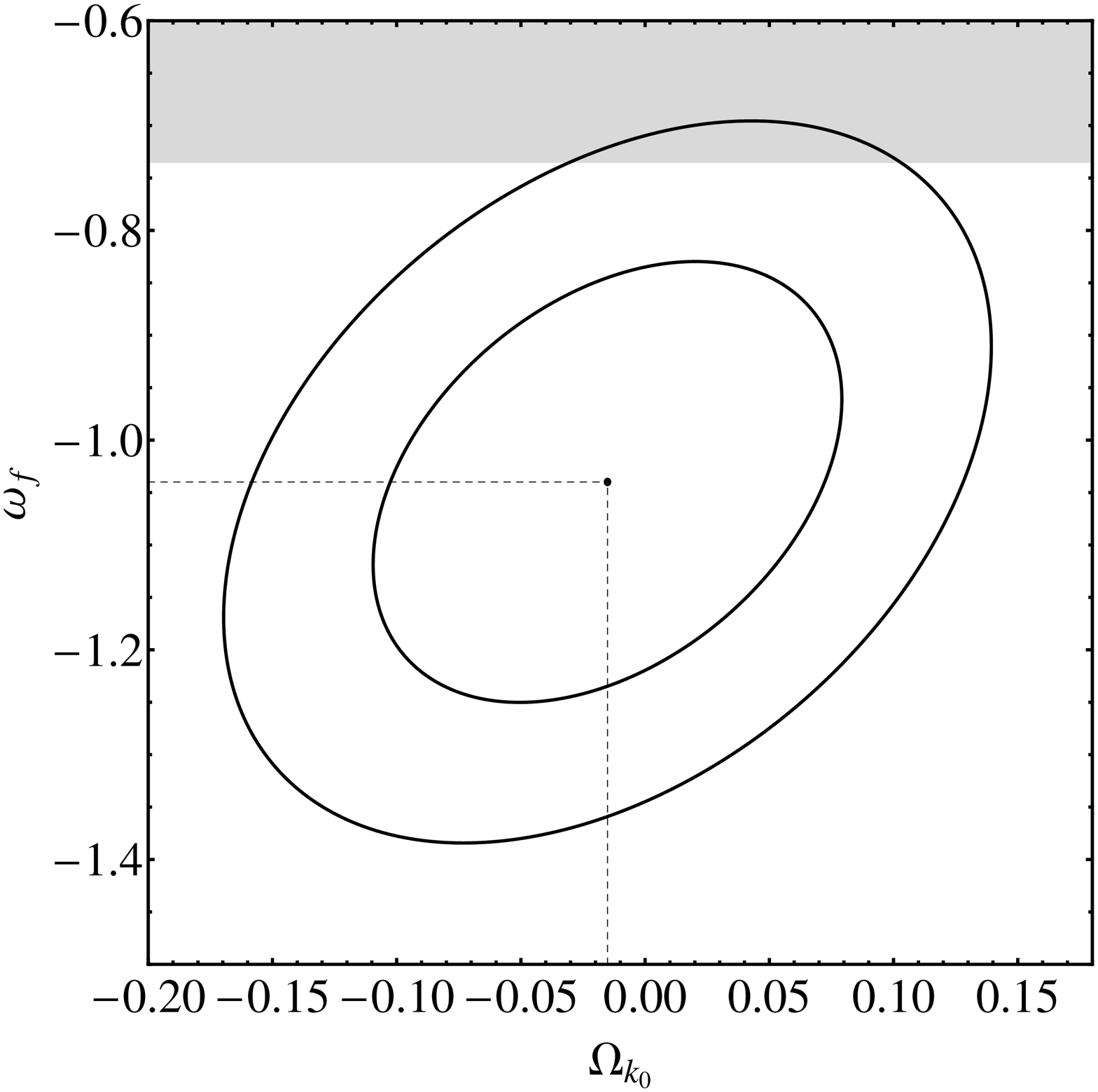}
\caption{\label{SnHz1} Contour plots (1$\sigma$ and 2$\sigma$) in the $\omega_0-\omega_f$ and $\Omega_{k_0}-\omega_f$ plane  for the joint constraint SnIa+H(z). The best fit parameters are indicated with dashed lines. The gray region is excluded given the condition $\omega_f<\omega_0$. The left panel is for the flat case and in the right panel we have chosen $w_0=-0.736$ as a prior.}
\end{figure}

\begin{table}[ht!]
\begin{tabular}{l l l l l}
\hline\hline
Dataset &\ \ \ \  $\chi^2_{min}$ &\ \ \ \ $\Omega_{k_0}$ &\ \ \ \ $\omega_0$ &\ \ \ \ $\omega_f$\\ 
\hline\hline
SnIa+H(z) &\ \ \ \ 570.974 &\ \ \ \ 0 (prior)&\ \ \ \ -0.736$\pm$0.061 &\ \ \ \ -1.038$\pm$0.193 \\
\hline
SnIa+H(z) &\ \ \ \ 570.963 &\ \ \ \ -0.015$\pm$0.139 &\ \ \ \ -0.736 (prior) &\ \ \ \  -1.040$\pm$0.062\\
\hline\hline
\end{tabular}
\caption{\label{Tf} The table shows the best fit parameters with the 1$\sigma$ uncertainty. We have performed Bayesian analysis with two free parameters and different prior in each case. In both cases we have imposed $\omega_0>\omega_f$.}
\end{table}

We also perform statistical analysis using the Hubble expansion rate data \cite{Hz}. In the same way as it was done with the $\tilde{\mu}_0$ parameter, we note that $H_0$ is a nuisance paremeter and, instead of minimize the function:
\beq
\chi_H^2(H_0,{\bf p})=\sum_{i=1}^{14}\left(\frac{H_{obs,i}-H_{th}(z_i;H_0;{\bf p})}{\sigma_H(z_i)}\right)^2,\ \ \ \textrm{where}\ H_{th}=H_0f({\bf p})
\eeq
we minimize the function $\tilde{\chi}_H^2$, which is independent of $H_0$,
\beq
\tilde{\chi}_{H}^2({\bf p})=A({\bf p})-\frac{B({\bf p})^2}{C({\bf p})}, \ \textrm{where now}
\eeq
\beq
A({\bf p})=\sum_{i=1}^{14}\frac{H_{obs,i}^2}{\sigma_{H}(z_i)^2};\ 
B({\bf p})=\sum_{i=1}^{14}\frac{H_{obs,i}f_i({\bf p})}{\sigma_{H}(z_i)^2};\
C({\bf p})=\sum_{i=1}^{14}\frac{f_i^2({\bf p})}{\sigma_{H}(z_i)^2}
\eeq
A joint analysis using SnIa+H(z) lead us to the best fit parameter showed in TABLE \ref{Tf} and FIG.\ref{SnHz1}. 

\section{Conclusion}
\label{cl}
We explore an alternative scheme for the problem of the dark sectors in Cosmology where dark energy and dark matter are described by the evolution of a single fluid. 
In particular, we analyse extensions to the DDE model proposed in \cite{Lim} by including spatial curvature and more general potentials.

We have found a family of potentials for which the model can be described by a three-dimensional autonomous system. We study the corresponding critical points and their characteristics. In general, there are two critical points which can be interpreted as the initial and final state of our universe. Namely, in the initial state the fluid behaved like dust whereas in the final state the fluid have a constant state parameter with $\omega_f<-\frac{1}{3}$. These results are independent of the spatial curvature.

In order to constraint the parameters of the model by using Bayesian analysis, we find numerical solutions to the set of Eqs.(\ref{eq6})-(\ref{eq7}). We found that the curvature has a negligible incidence in the evolution of the state parameter $\omega$, as it is shown in FIG. \ref{mono2}.
 
The Bayesian analysis shows that this model is consistent with the available data from SnIa and H(z). For null spatial curvature the best fit values for the parameters are $\omega_0=-0.736\pm0.061$ and $\omega_f=-1.038\pm0.193$, which suggest a final state dominated by a phantom-type fluid and a crossing of the so called phantom barrier in the future evolution. We note that the value of $w_0$ is consistent with the best fit results for $\Lambda $CDM when a single effective fluid is considered \cite{WMAP7}.

When spatial curvature is taken into account (choosing as prior $\omega_0=-0.736$), the best fit values are $\Omega_{k_0}=-0.015\pm0.139$ and $\omega_f=-1.040\pm0.062$, slightly favouring a closed model.
In both cases (with and without curvature), the best fit value for $\omega_f$ is consistent with a nearly flat scalar potential.

In the near future we expect to study cosmological perturbations on this model in order to explore possible deviations of the standard picture and include more data as WMAP and BAO.

\begin{acknowledgments}
This work has been partially supported by Universidad del B\'io-B\'io through grant DIUBB 121407 GI/VC.
A.C. is supported by Fondecyt grant N$^\circ$ 11110507. P. L. is supported by Fondecyt grant N$^\circ$ 11090410. 
\end{acknowledgments}

\end{document}